\documentclass[fleqn,usenatbib,usedcolumn]{mnras}

\usepackage{newtxtext,newtxmath}
\renewcommand{\la}{\lesssim} % for less than similar from newtxmath, not \la from mnras.cls
\renewcommand{\ga}{\gtrsim} % for greater than similar from newtxmath, not \la from mnras.cls
\usepackage[T1]{fontenc}
\usepackage{graphicx}	% Including figure files
\usepackage{amsmath}	% Advanced maths commands
\usepackage{amssymb}	% Extra maths symbols
\usepackage[export]{adjustbox}
\usepackage[usenames,dvipsnames]{color}

\newcommand{\Sref}[1]{Section \ref{#1}}
\newcommand{\Tref}[1]{Table \ref{#1}}

\sfcode`\.=1001\sfcode`\?=1001\sfcode`\!=1001
\newcommand{\Fref}[1]{\ifhmode \ifnum\spacefactor=1001 Figure \ref{#1}\else Fig.\ \ref{#1}\fi \else Figure \ref{#1}\fi}
\newcommand{\Eref}[1]{\ifhmode \ifnum\spacefactor=1001 Equation (\ref{#1})\else equation (\ref{#1})\fi \else Equation (\ref{#1})\fi}

\newcommand{\cmm}{cm$^{-2}$}

\newcommand{\kms}{\ensuremath{\textrm{km\,s}^{-1}}}

\newcommand{\lya}{\ensuremath{\textrm{Ly}\alpha}}

\newcommand{\zem}{\ensuremath{z_\textrm{\scriptsize em}}}
\newcommand{\zab}{\ensuremath{z_\textrm{\scriptsize abs}}}
\newcommand{\mNHI}{N_\text{H\kern 0.2em\textsc{i}}}
\newcommand{\NHI}{\ensuremath{N_\textsc{h\scriptsize{\,i}}}}
\newcommand{\nH}{\ensuremath{n_\textsc{h}}}
\newcommand{\NH}{\ensuremath{N_\textsc{h}}}

\newcommand{\lNHI}{\ensuremath{\log_{10}(N_\textsc{h\scriptsize{\,i}}/\textrm{cm}^{-2})}}

\newcommand{\lnH}{\ensuremath{\log_{10}(n_\textsc{h}/\textrm{cm}^{-3})}}
\newcommand{\tran}[3]{\ensuremath{\ion{#1}{#2}\,\lambda\textrm{#3}}}
\newcommand{\MH}[1]{\ensuremath{<ft[\textrm{M}/\textrm{H}\right]}}

\newcommand{\popler}{\ensuremath{\textsc{uves\_popler}}}
\newcommand{\HI}   {{\rm H}{\sc \,i}}
\newcommand{\msun}{$M_\odot$}

\newcommand{\lmetal}{\ensuremath{\log_{10} (Z / Z_\odot)}}

\newcommand{\nHI}{$n_\mathrm{H\kern  0.1em \scriptsize \textsc{i}}$}
\newcommand{\mnHI}{n_\mathrm{H\kern 0.1em\scriptsize \textsc{i}}}

\newcommand{\CII}  {{\rm C}{\sc \,ii}}

\newcommand{\CIV}  {{\rm C}{\sc \,iv}}

\newcommand{\SiII} {{\rm Si}{\sc \,ii}}

\newcommand{\SiIV} {{\rm Si}{\sc \,iv}}

\usepackage{array}
\newcolumntype{:}{>{\global<t\currentrowstyle\relax}}
\newcolumntype{;}{>{\currentrowstyle}}

\title[Apparently metal-free gas at $z=4.4$]{Exploring the origins of a new, apparently metal-free gas cloud at \boldmath{$z=4.4$}}

\author[P. Fr\'{e}d\'{e}ric Robert et al.]{P. Fr\'{e}d\'{e}ric Robert$^{1}$\thanks{E-mail: probert@swin.edu.au (P. Fr\'{e}d\'{e}ric Robert)}, Michael T. Murphy,$^1$ John M. O'Meara,$^2$ Neil H. M. Crighton,$^1$\newauthor Michele Fumagalli$^3$ \\
  $^{1}$Centre for Astrophysics and Supercomputing, Swinburne University of Technology, Hawthorn, Victoria 3122, Australia\\
  $^{2}$Department of Chemistry \& Physics, Saint Michael's  College, One Winooski Park, Colchester VT, 05439\\
  $^{3}$Institute for Computational Cosmology and Centre for Extragalactic Astronomy, Department of Physics, Durham University, South Road,\\ Durham, DH1 3LE, United Kingdom\\
}

\date{Accepted ---. Received ---; in original form ---}

\pubyear{2018}

\begin{document}
\label{firstpage}
\pagerange{\pageref{firstpage}--\pageref{lastpage}}
\maketitle
% Single paragraph, not more than 250 words (200 for Letters), no references.
\begin{abstract}
We report the discovery and analysis of only the third Lyman-limit system in which a high-quality resolution, echelle spectrum reveals no metal absorption lines, implying a metallicity $\la$1/10000 solar. Our HIRES spectrum of the background quasar, PSS1723$+$2243, provides a neutral hydrogen column density range for LLS1723 of $\NHI=10^{\text{17.9--18.3}}$\,cm$^{-2}$ at redshift $\zab\approx4.391$. The lower bound on this range, and the lack of detectable absorption from the strongest low-ionisation metal lines, are combined in photoionisation models to infer a robust, conservative upper limit on the metallicity: $\log(Z/Z_\odot)<-4.14$ at 95\% confidence. Such a low metallicity raises the question of LLS1723's origin and enrichment history. Previous simulations of the circumgalactic medium imply that LLS1723 is a natural candidate for a cold gas stream accreting towards a galaxy. Alternatively, LLS1723 may represent a high-density portion of the intergalactic medium containing either pristine gas -- unpolluted by stellar debris for 1.4\,Gyr after the Big Bang -- or the remnants of low-energy supernovae from (likely low-mass) Population III stars. Evidence for the circumgalactic scenario could be obtained by mapping the environment around LLS1723 with optical integral-field spectroscopy. The intergalactic possibilities highlight the need for -- and opportunity to test -- simulations of the frequency with which such high-density, very low-metallicity systems arise in the intergalactic medium. 
\end{abstract}


%%%%%%%%%%%%%%%%%%%%%%%%%%%%%%%%%%%%%%%%%%%%%%%%%%

%%%%%%%%%%%%%%%%% BODY OF PAPER %%%%%%%%%%%%%%%%%%

\section{Introduction}\label{s:intro}

Galaxies form and evolve through interactions with their gaseous environments: the circumgalactic medium (CGM) and the intergalactic medium (IGM)\footnote{Here we consider the CGM as the environment within one virial radius of a galaxy, and the IGM as that beyond a virial radius permeating the space between galaxies.}. The CGM and IGM bear the imprints of several astrophysical processes such as the initial formation of Population III (PopIII) stars in the proto-galaxies, accretion of enriched gas which fuels the formation of PopII/I stars inside the galaxies, and expulsion of gas due to the death of these stars as supernovae. All these processes are sources of contamination with metals. Therefore, gas ``clouds'' found in these environments (with large neutral hydrogen column densities, $\NHI\ga10^{17.2}$\,\cmm) should, in general, be expected to have metallicities that reflect such processes. However, using quasar absorption spectroscopy, two gas clouds with no apparent associated metal lines were discovered serendipitously at $z \sim 3$ by \citet{2011Sci...334.1245F}, with upper limits for their metallicity of $\lmetal \la -4$.

These two ``clouds'' are Lyman limit systems (LLSs): neutral hydrogen gas column densities of $17.2 < \lNHI < 20.3$ so that they are optically thick bluewards of the Lyman limit ($\lesssim 912$\,\AA). The physical properties of LLSs have been investigated in several surveys by differents groups at $z \lesssim 1$ \citep[e.g.][]{2013ApJ...770..138L,2016ApJ...833..283L,2016ApJ...831...95W} and $z \gtrsim 2$ \citep[e.g.][]{1990ApJS...74...37S,2013ApJ...775...78F,2015ApJS..221....2P,2015ApJ...812...58C,2016MNRAS.455.4100F,2016ApJ...833..283L}. In particularly, the metallicity distribution of LLSs at low redshift spans the range $-2.0 \la \lmetal \la 0.4$ and may be bimodal, with low and high metallicity branches at $\lmetal \sim -1.6$ and $-0.3$ \citep{2013ApJ...770..138L}. At higher redshift, which is the focus of this paper, LLSs exhibit an even broader metallicity distribution, spanning over 4 magnitudes \citep{2016MNRAS.455.4100F}. This distribution shows that high-redshift LLSs are in general metal-poor, with a peak at $\lmetal \approx -2$, but that very metal-poor LLSs, with $\lmetal < -3$, are relatively rare. Indeed, only two LLSs have been reported to have metallicities below this value -- $\lmetal \approx -3.4$ -- which are derived from secure metal-line detections \citep{2016MNRAS.457L..44C,2016ApJ...833..283L}. These systems were identified in large samples of existing high-resolution ($R\sim50000$) quasar spectra from the HIRES spectrograph \citep{1994SPIE.2198..362V} on the Keck I telescope \citep{2015AJ....150..111O} and UVES \citep{2000SPIE.4008..534D} on the UT2 Very Large Telescope (VLT) \citep{2019MNRAS.482.3458M}. The very low metallicities of these LLSs, and the two apparently metal free systems of \citet{2011Sci...334.1245F} raises the question of their origins.

Higher metallicity LLSs (i.e.\ $\lmetal > -3$) are normally considered to originate in the CGM \citep[e.g.][]{1989ApJS...69..703S,2016ApJ...833..283L,2016MNRAS.455.4100F,2016MNRAS.455.4100F,2016ApJ...831...95W}. In this picture, very low metallicity LLSs would potentially represent streams of relatively unenriched cold gas being accreted into the circumgalactic environment \citep[e.g.][]{2005MNRAS.363....2K,2006MNRAS.368....2D,2009Natur.457..451D,2011MNRAS.418.1796F,2011MNRAS.412L.118F,2012MNRAS.421.2809V}. Nevertheless, an alternative possibility is that at least some very low metallicity LLSs arise in the IGM. This scenario was recently explored as the possible origin for the lowest metallicity LLS with metal-line detections by \citet{2016MNRAS.457L..44C}. This ``near-pristine'' absorber, and the next lowest metallicity LLS reported by \citet{2016ApJ...833..283L}, have carbon-to-silicon ratios consistent with predictions of nucleosynthesis models of PopIII stars, though they are also consistent with some PopII models \citep{2002ApJ...567..532H}. The possible PopIII origin, in which metal contamination occurred at $z\ga10$ and no further pollution occurred for at least another 1--2 billion years, implies that the LLSs are likely in the IGM, not in close proximity to a galaxy. The photoionisation model of \citeauthor{2016MNRAS.457L..44C}'s LLS also implies a large size, albeit with a large possible range, tentatively supporting the idea that this cloud arises in the IGM and not the CGM. Attempting to distinguish between these two scenarios will require larger samples of very low metallicity absorbers, which motivates targeted searches for them instead of relying on further serendipitous discoveries or large samples of existing high-resolution quasar spectra.

In this paper we present the analysis of LLS1723, a new, apparently metal-free LLS, with a conservative metallicity upper limit of $\lmetal < -4.14$ (95\%-confidence), and discuss its possible origins. In \Sref{s:observations}, we describe how we conducted a dedicated search, specifically targeting very metal-poor LLSs, in a Keck observing campaign using the HIRES spectrograph. In \Sref{s:analysis}, we detail the analysis of the absorption line features of LLS1723, allowing us to establish a fiducial model for its hydrogen and metal content that provides the most conservative metallicity upper limit stated above. \Sref{s:results} explains our photoionisation modelling of LLS1723, which provides the metallicity estimate, and a series of consistency checks we conducted to test its robustness. Finally, \Sref{s:discussion} puts this very low metallicity result into context with a discussion of the possible origins of LLS1723, distinguishing between the CGM and IGM scenarios outlined above. Our main conclusions are summarised in \Sref{s:conclusion}.

\section{Target selection, observations, and data reduction}\label{s:observations}

As stated in \Sref{s:intro}, our goal was to conduct a dedicated search of very metal-poor LLSs, and to study their properties and understand their possible origins. Our initial sample was composed of 10 LLSs. Two are from the sample of \citet{2015ApJ...812...58C} and 8 are from the survey of \citet{2015ApJS..221....2P}:
\begin{itemize}
\item \citet{2015ApJ...812...58C} analysed and provided metallicity estimates for 17 LLSs at $z>2$ observed with the MagE spectrograph \citep{2008SPIE.7014E..54M} on the Magellan Clay telescope. This was a sub-sample drawn from the 96 LLSs with no detected metal lines in the full sample of 194 found by \citet{2010ApJ...718..392P} in Data Release 7 \citep{2009ApJS..182..543A} of the Sloan Digital Sky Survey (SDSS) quasar spectra. This full sample of 194 LLSs should not contain strong metallicity biases; however, given the sample selections above, the effective unbiased sample size is $\approx17/96\times194=34$ LLSs. From the sample of 17 LLSs we selected two because their metallicity upper limit were found to be $\lmetal < -3$ by \citet{2015ApJ...812...58C}.
\item For the 157 LLSs found in the survey of \citet{2015ApJS..221....2P}, \citet{2016MNRAS.455.4100F} determined the metallicity for each LLS, so this sample should not contain strong metallicity biases. We selected 8 LLSs with metallicity estimates $\lmetal\la-3$: two of these 8 had existing high-resolution spectra available in which metal lines were detected, which allowed \citet{2016MNRAS.455.4100F} to infer the metallicity; the other 6 only had lower-resolution spectra obtained with ESI \citep{2002PASP..114..851S} on the Keck II telescope or MIKE \citep{2003SPIE.4841.1694B} on the Magellan Clay telescope, in which no metal lines were detected and, therefore, only metallicity upper limits could be derived.
\end{itemize}
We targetted the 8 LLSs without existing high-resolution spectra (2 from \citealt{2015ApJ...812...58C} and 6 from \citealt{2015ApJS..221....2P}) with HIRES during three observing runs in 2016--2017, obtaining signal-to-noise ratios $\ga$60 per $\approx$2.3\,\kms\ pixel in the continuum near the expected wavelength of the strongest metal absorption lines. One of these LLSs, LLS1723 appeared to be free of metal lines in the HIRES spectra and indicated a significantly lower metallicity than the others, so we report our analysis of this system here. The remaining systems will be analysed and reported in future papers.

The very metal-poor absorption system, LLS1723, was first identified in \citet{2015ApJS..221....2P}, towards the $\zem=4.515$ quasar PSS1723$+$2243 (hereafter J1723$+$2243), based on absorption features at redshift $\zab=4.391$, using an ESI spectrum. At the estimated absorber redshift, the strongest metal absorption lines of \CII, \CIV, \SiII\ and \SiIV\ were not detected in the ESI spectrum. Using this same spectrum, the physical properties of LLS1723 were studied by \citet{2016MNRAS.455.4100F}, among which the posterior probability distribution function for the metallicity of the absorbing gas was derived using photoionisation models. With access to these estimates of LLS1723's neutral hydrogen column density (\NHI), metallicity and ionic column densities, we chose to target J1723$+$2243 as part of our search of very metal-poor systems with HIRES.

We observed J1723$+$2243 (right ascension $17^{\rm h}23^{\rm m}23.2^{\rm s}$, declination $+22^{\circ}43'58"$ in the J2000 epoch) with HIRES on 2017 UT June 13, for a total exposure time of 20665 seconds, obtained with 6 exposures. HIRES was configured with the red cross-disperser, and we chose a slit width of 1.148" (C5 decker) to provide a resolving power of $R=37500$. The 6 exposures were taken with two wavelength settings which, when combined, cover with gaps 4786 to 9217 \AA\ with small ($\sim$2--30\,\AA) gaps between echelle orders at wavelengths beyond 6922\,\AA, and two larger gaps ($\approx$80\,\AA\ at 6168 and 7764\,\AA\ respectively) between the three CCDs in the HIRES detector. The initial reduction steps performed with the {\sc makee} data reduction pipeline\footnote{See \urlstyle{rm}\url{http://www.astro.caltech.edu/~tb/makee/}.}. To locate the spatial center of each echelle order, trace exposures were obtained using a flat-field observation though a pinhole decker (D5). Flat-field exposures, observed  through the science slit, were used to correct for pixel-to-pixel sensitivity variations and for the blaze function. Each quasar exposure was followed by a thorium--argon (ThAr) lamp exposure without changing the spectrograph set-up. {\sc makee} establishes the wavelength scale for each echelle order of a quasar exposure using a sixth-order polynomial fit to the ThAr line wavelengths versus their detector position along the order. However, for the reddest orders, {\sc makee} failed to do so: not enough lines were identified on the ThAr exposures in {\sc makee}'s library of reference wavelength solutions for different settings. Because some of the key metal-line transitions for LLS1723 were expected in these reddest orders, we manually identified ThAr lines, and iteratively built up new reference solutions for the exposures in our particular wavelength settings. The final wavelength solutions had residuals with root-mean-square (RMS) deviations from the mean of less than $0.2$ pixels ($\sim 0.5$\,\kms), and {\sc makee} was able to provide a 1D extracted spectrum for each exposure.

\begin{figure*}
  \begin{center}
  \includegraphics[width=0.95\textwidth,valign=t]{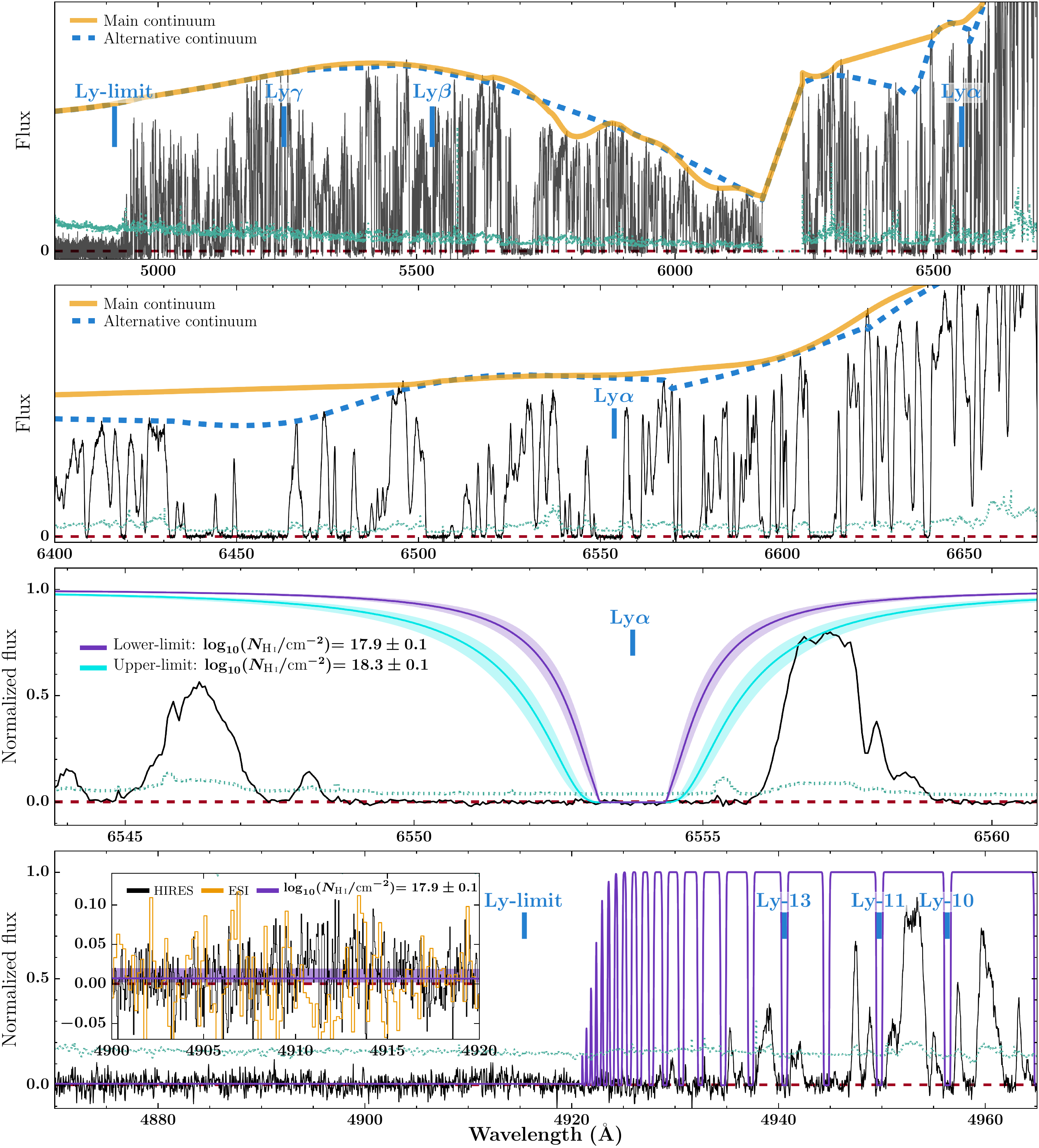}
  \vspace{-0.2cm}
  \caption{Overview of the hydrogen absorption from LLS1723. First panel: HIRES quasar spectrum of J1723$+$2243, with our two possible continua superimposed: the main continuum used for our fiducial analysis (yellow solid line) and the alternative continuum (blue dashed line) we use as a consistency check. Some Lyman series absorption lines from LLS1723 are marked with blue vertical ticks and labels. Second panel: Smaller region around the \lya\ absorption line of LLS1723 illustrating the two different continua in more detail. Third panel: \lya\ absorption line for LLS1723 in the HIRES spectrum of J1723$+$2243, normalized by the main continuum of the previous two panels. Two possible \HI\ models for LLS1723 are represented: the lower limit derived from the Lyman limit (lowest panel) shown in purple, and the upper limit derived from the \lya\ line in this panel shown in cyan. The shading around each model indicates the systematic $0.1$\,dex column density uncertainty. Both share the same redshift, $\zab=4.381085$, and Doppler broadening parameter, $b=8$\,\kms. We use the lower limit on \HI\ to derive a conservative upper limit for the metallicity of LLS1723. Fourth panel: Lyman limit for LLS1723 in the HIRES spectrum, normalized by the main continuum. The inset highlights the lack of detected flux below the Lyman limit in the HIRES spectrum (black histogram) and ESI spectrum (orange histogram). The lower limit \HI\ model is shown by the purple solid line in the main panel and inset, with the shading corresponding to its systematic $0.1$\,dex column density uncertainty. In all panels, the zero level is indicated by red dashed lines, and the 5$\sigma$ flux uncertainty array is indicated with a green dotted line.}
   \label{f:grid}
    \vspace{-0.4cm}
  \end{center}
\end{figure*}

The extracted quasar exposures were combined to form the final spectrum using \popler\ \citep{Murphy:2016:UVESpopler} with the same approach as used in \citet{2019MNRAS.482.3458M}. Briefly, each echelle order of each exposure was redispersed onto a common wavelength scale converted to vacuum in the heliocentric reference frame, with a dispersion of 2.3\,\kms\ per pixel. For each echelle order, the \popler\ code automatically scales the flux and error arrays so that the flux optimally matches that in overlapping orders from all exposures. However, in regions of very low, or zero flux, this approach fails. Therefore, we manually scaled the flux arrays of the echelle orders falling just below the Lyman limit to match the relative scaling of the echelle orders just above it in their corresponding exposures. This is important because, as described in \Sref{s:hydro}, the lack of detected flux below the Lyman limit determines our lower limit on \NHI\ for LLS1723 -- see bottom panel of \Fref{f:grid}. As a consistency check, we used the ESI spectrum of J1723$+$2243, which did not show evidence of remaining flux for the same region -- see \Sref{s:hydro} and inset of bottom panel in \Fref{f:grid}. The scaled flux arrays of all exposures are combined by \popler, with their inverse variances acting as weights, in an iterative way to remove deviant pixels.

\popler\ automatically estimates a first-guess continuum using a polynomial fitting approach.  Although this continuum was reasonable for the region redwards of the \lya\ emission line of J1723$+$2243, this was not the case for the \lya\ forest region, particularly the \lya\ and Lyman limit regions of LLS1723. This is a typical feature for quasar spectra at $z \sim 4$\ given the large opacity of the \lya\ forest. Therefore, we manually set the continuum for these regions. In general, we selected very narrow wavelength windows showing the least \lya\ forest absorption and connected them with the polynomial continuum fitting feature of \popler. However, for the region around the \lya\ line of LLS1723, depicted in the top two panels of \Fref{f:grid}, this approach was unstable due to the lack of enough such peaks in the spectrum. We therefore manually selected node and a spline function to fit a continuum around the \lya\ line of LLS1723. However, this is still anchored to the observed flux in two places and so, given the expected high number-density of the \lya\ forest at $z\approx4.4$, we are still likely underestimating the continuum level. As detailed in \Sref{s:hydro}, this ensures that even our upper bound \NHI\ for LLS1723 provides a conservative metallicity upper limit. Clearly, this manually-set continuum will have a considerable human dependence. To help gauge this variance, two continua were created by two different authors: a ``main continuum'' from which all results in this paper are derived, and an ``alternative continuum'' which we use to check the variance in our main results; these are depicted as solid orange and dashed blue lines top two panels of \Fref{f:grid}. Despite the clear differences, the metallicity upper limit we derive for LLS1723 does not change significantly using the alternative continuum, as described in \Sref{s:results}. The continuum-to-noise ratio of the final spectrum is 31 per 2.3\,\kms\ pixel at LLS1723's Lyman limit, 58 per pixel at its \lya\ line, and 63 per pixel at the most important metal-line transition, \tran{Si}{ii}{1260}.

\section{Analysis}\label{s:analysis}

\subsection{Hydrogen lines analysis}\label{s:hydro}

\begin{figure}
  \begin{center}
    \includegraphics[width=0.95\columnwidth]{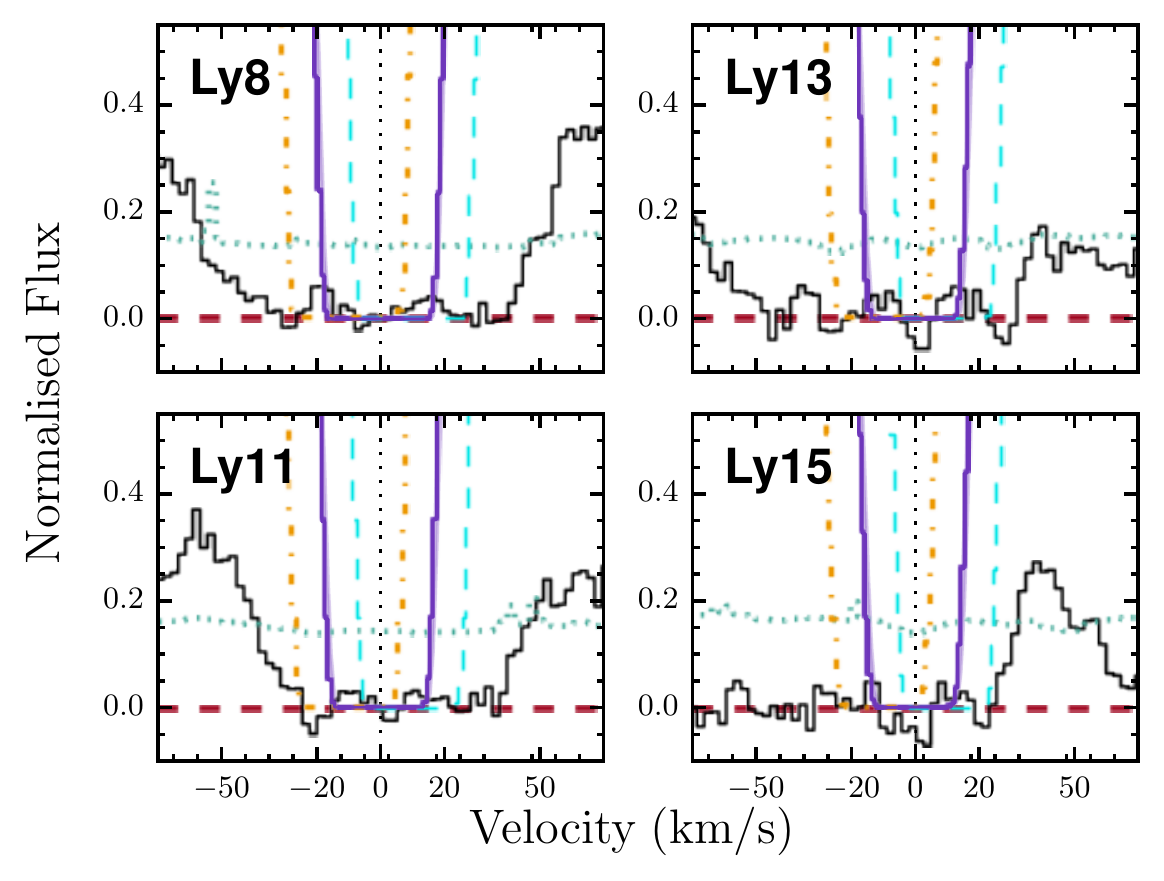}
    \vspace{-0.3cm}
    \caption{Normalized flux for the Lyman series lines used for constraining the combination of redshift and Doppler $b$ parameter of LLS1723. The zero velocity redshift is set at $\zab=4.391085$. The purple solid lines correspond to our fiducial model of LLS1723 with $\lNHI=17.9 \pm 0.1$, $\zab=4.391085$ and $b=8$\,\kms; the line thickness represents the systematic uncertainty on \lNHI. The orange dash-dotted lines and cyan dashed lines correspond respectively to the models with $z_\textrm{\scriptsize blue}=4.390905$, and $z_\textrm{\scriptsize red}=4.391265$; both with $\lNHI=17.9 \pm 0.1$\ and $b=8$\,\kms. These two models illustrate the uncertainty in redshift of LLS1723 by shifting \zab\ by $\pm$10\,\kms. In all panels, the zero level is indicated by red dashed lines, and the 5$\sigma$ flux uncertainty array is indicated with a green dotted line.}
    \label{f:lymanseries}
    \vspace{-0.5cm}
  \end{center}
\end{figure}

We modelled the Lyman series of LLS1723 using Voigt profiles with custom {\sc python} visual tools. Our model is a single phase gas cloud characterised by three parameters: the absorption redshift \zab, the neutral hydrogen column density \NHI, and the Doppler broadening parameter $b$. Using an ESI spectrum, \citet{2015ApJS..221....2P} derived initial estimates for these parameters: $\zab=4.391$, $\lNHI=18.25 \pm 0.25$, and $b=30$\,\kms. At $\zab=4.391$, the \lya\ forest is very thick, so many Lyman series lines were checked in order to constrain the position (\zab) and width of the line ($b$), guided by a model with the initial parameter estimates from \citet{2015ApJS..221....2P} overlayed on our HIRES spectrum. The Lyman transitions that most constrain \zab\ and $b$ are Ly8, Ly11, Ly13, and Ly15 as depicted in \Fref{f:lymanseries}. We adjusted the values of $\zab$ and $b$ until they best matched the small rises in flux at $\pm$10--15\,\kms, seen most clearly in Ly8 and Ly13. This restricted the absorption redshift to $\zab=4.391085$ and a maximum Doppler parameter of $b\approx8$\,\kms. However, it is clear in \Fref{f:lymanseries} that the small flux rises at $\pm$10--15\,\kms\ may not be significant, and that a slightly larger range of redshifts and/or $b$ parameters may be allowed. The steep flux rises at $\pm$20\,\kms\ seen in Ly11 and Ly15 in \Fref{f:lymanseries} restrict the combined change in \zab\ and $b$. We parametrise this uncertainty by shifting the absorber by $\pm$10\,\kms, to $z_\textrm{\scriptsize blue}=4.390905$ and $z_\textrm{\scriptsize red}=4.391265$, and reassessing our \NHI\ measurements below. The alternative approach, of increasing the $b$-parameter to $\approx$20\,\kms, has no effect on the lower limit on \NHI\ or the metal line column densities derived below, and so does not affect the associated metallicity estimate appreciably.

Having established the redshift of LLS1723, its \NHI\ value can be assessed. However, precisely estimating \NHI\ for LLSs with $\lNHI \la 19$ is usually challenging because the \lya\ line does not have strong damping wings (it is on the flat part of the curve of growth) and the Lyman series lines are saturated. These problems are evident in Figs.\ \ref{f:grid} and \ref{f:lymanseries}. Instead, we have used the Lyman limit and the \lya\ line to provide a plausible range of values for \NHI, corresponding to a lower and an upper limit respectively, of $\lNHI=17.9 \pm 0.1$ and $18.3 \pm 0.1$. We discuss these in turn below.

For our main result -- that LLS1723 has a very low metallicity -- the lower limit on \NHI\ is most important to establish robustly as it gives the highest possible metallicity upper limit. The lack of significant flux observed below the Lyman limit provides a strict lower-limit on \NHI\ for LLS1723, as shown in the bottom panel of \Fref{f:grid}. This will be insensitive to the $b$-parameter and also the precise value of the redshift within the $\pm$10\,\kms\ range discussed above. Using the redshift and $b$ parameter established above (4.391085 and 8\,\kms), we adjusted the \NHI\ value until the model was minimally consistent with the lack of flux at 4900--4920\,\AA\ just below the Lyman limit. The inset in the bottom panel of \Fref{f:grid} shows this most clearly. The lack of flux was further established using the lower-resolution ESI spectrum in the same wavelength range, also depicted in the bottom panel of \Fref{f:grid}. An uncertainty of 0.1\,dex in \lNHI\ was determined visually for the \NHI\ lower limit to approximately match the scatter in the HIRES and ESI flux values and so that their mean clearly fell below the lowest allowed value [i.e.\ $\lNHI=17.8$], as shown by the shaded region of \Fref{f:grid}'s bottom panel. The top and bottom panels of \Fref{f:grid} also show that the continuum placement at the Lyman limit is uncertain by as much as $\pm$15\%. However, changing the continuum level by this amount required an adjustment of the \NHI\ lower limit by $<$0.05\,dex, well within the 0.1\,dex uncertainty range already established. Finally, we searched the HIRES spectrum of J1723$+$2243 for other LLSs that may cause total absorption below the Lyman limit of LLS1723, but we did not find any plausible candidates within $\sim$30\,\AA. That is, the lack of flux observed below LLS1723's Lyman limit appears to be due only to LLS1723 and not other unrelated LLSs at a nearby redshift: our \NHI\ lower limit of $\lNHI=17.9 \pm 0.1$ is robust.

We obtained the upper limit, $\lNHI=18.3\pm 0.1$, based on the observed transmission near the \lya\ line of LLS1723. As shown on the third panel of \Fref{f:grid}, the flux peak at 6557\,\AA\ limits the total absorption allowed in the model at that point. We increased the \NHI\ value until that constraint was satisfied, as shown by the cyan line and shading in \Fref{f:grid}. The figure also shows that the \NHI\ lower limit model is constrained by the 6557\,\AA\ flux peak. Like the \NHI\ lower limit, the uncertainty we assign to the upper limit is systematic in nature but, in this case, is dominated by the uncertainty on the continuum placement. As described in \Sref{s:observations}, we fitted two different continua to the \lya\ region of the HIRES spectrum, plotted as the ``main'' and ``alternative'' continua in \Fref{f:grid}. These continua provided very similar \NHI\ upper limits, differing only by $\approx$0.05\,dex. However, our general approach to establishing these two continua were quite similar: in both cases, the continuum is an interpolation between the least absorbed flux peaks in the surrounding spectral regions. Given the high expected number-density of \lya\ forest lines at this redshift, this approach almost certainly underestimates the continuum level. Increasing the main continuum level by a 10\%, which we regard as conservative, requires the \lNHI\ upper limit to be increased to $18.4$. Therefore, we ascribe an uncertainty of 0.1\,dex to the \NHI\ upper limit.

As discussed above, we must reassess these lower and upper limits on \NHI\ for the $\pm$10\,\kms\ uncertainty in the \HI\ redshift. The lower limit on \NHI\ is insensitive to this uncertainty because it relies entirely on the lack of observed flux below the Lyman limit. However, shifting the upper limit model for \NHI\ by $\pm$10\,\kms\ changes the amount of absorption allowed at the 6557\,\AA\ flux peak: $\lNHI=18.35 \pm 0.1$ for $z_\textrm{\scriptsize blue}=4.390905$ and $\lNHI=18.20 \pm 0.1$ for $z_\textrm{\scriptsize red}=4.391265$. We test the effect of these changes on the metallicity results in \Sref{s:results} and find them to be very small; they also do not affect our main conclusions which are derived from the \NHI\ lower limit.

\subsection{Metal lines analysis}\label{s:metals}

\begin{figure}
  \begin{center}
    \includegraphics[width=0.95\columnwidth]{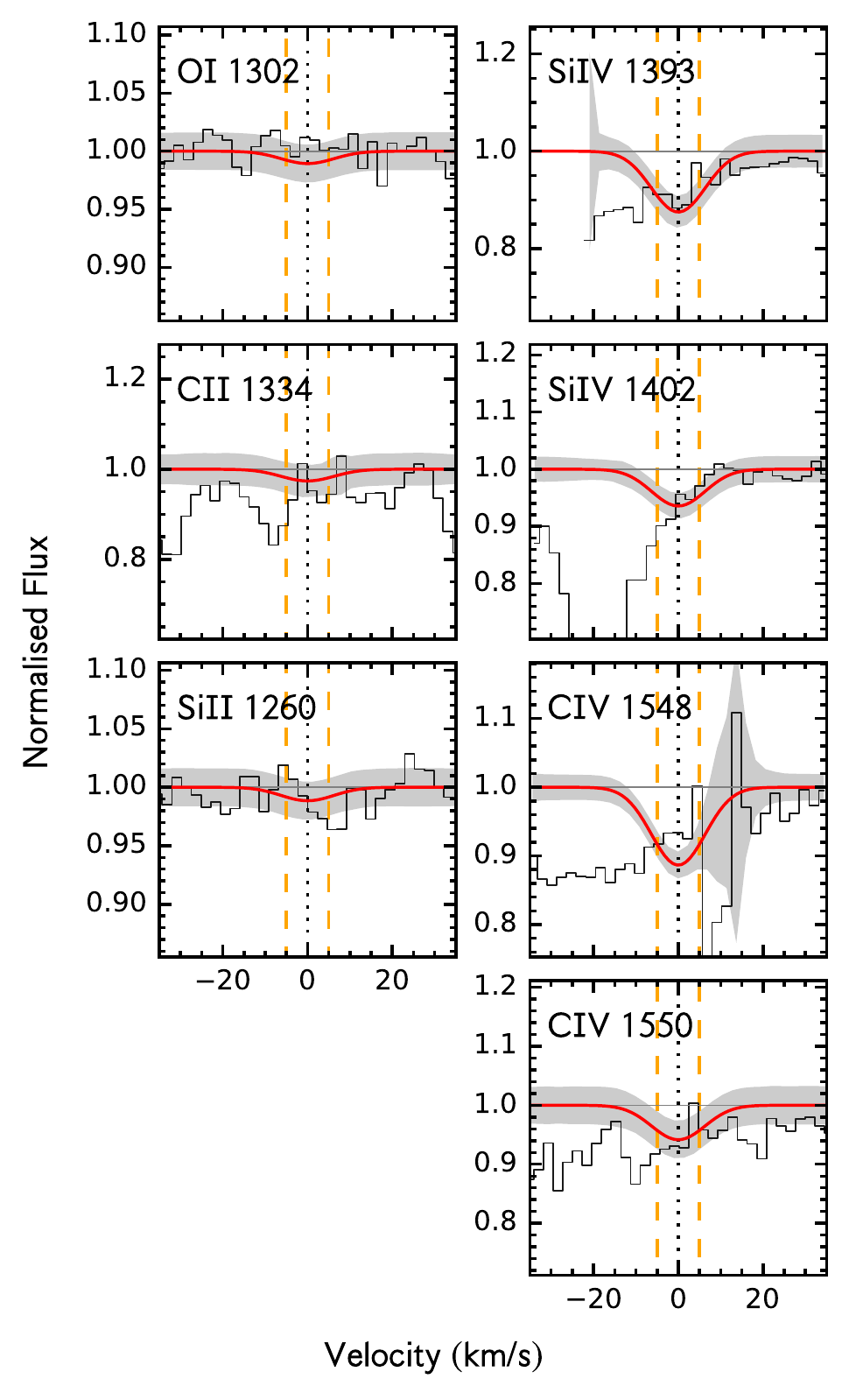}
    \vspace{-0.5cm}
    \caption{Regions of the J1723$+$2243 spectrum (black histogram) where the strongest metal absorption lines of LLS1723 are expected, centered at $\zab=4.391085$. The solid red lines are single-component Voigt profiles with $b=8$\,\kms\ at $\zab=4.391085$ with column densities set to the the 2$\sigma$ upper limits in the first column of \Tref{t:TEST}. The grey shading shows the 1$\sigma$ uncertainty in the flux. The vertical orange dashed lines indicate the velocity interval of $\pm$5\,\kms\ over which the column density upper limits were integrated using the apparent optical depth method. Note that for \ion{C}{iv}\ and \ion{Si}{iv} our approach is to obtain a 2$\sigma$ upper limit on the ionic column density by assuming that the total absorption seen at $v\approx0$\,\kms\ is real.}
    \label{f:metals}
    \vspace{-0.5cm}
  \end{center}
\end{figure}

A careful inspection of the HIRES spectrum of J1723$+$2243 did not reveal any metal absorption lines at the redshift of the \HI\ absorption ($\zab=4.391085$). \Fref{f:metals} depicts the strongest (highest oscillator strength) transitions of the most abundant metal species expected to be observed. No compelling detection is evident among these transitions. While the non-detections of \tran{O}{i}{1302} and \tran{Si}{ii}{1260} are clear, there are several weak, narrow features surrounding the \tran{C}{ii}{1334} transition apparent in \Fref{f:metals}. However, these are all attributable to telluric absorption lines and they are weakest within $\pm$5\,\kms\ of expected \tran{C}{ii}{1334} wavelength. For these three low-ion transitions, we determined 2$\sigma$ upper limits on their column densities using the apparent optical depth method of \citet{1991ApJ...379..245S}, with $\zab=4.391085$ and $b=8$\,\kms, integrated over the $\pm$5\kms\ velocity interval indicated on \Fref{f:metals}. These upper limits are listed in \Tref{t:TEST}.

\begin{table*}
\renewcommand{\arraystretch}{1.15}
\addtolength{\tabcolsep}{-3.6pt}
\footnotesize
\begin{center}
\caption{Metal column density upper limits (2$\sigma$), \NHI\ column density ranges and corresponding metallicity upper limits. We used the apparent optical depth method for the metal lines, integrated over a velocity interval of $\pm$5\,\kms, assuming optically thin transitions. The values listed in the second column correspond to the fiducial model used for our main-results. The last two columns correspond to the uncertainty in the redshift of LLS1723 introduced in \Sref{s:hydro}. The uncertainty was parametrised by shifting the absorber by $\pm$10\,\kms, to $z_\textrm{\scriptsize blue}=4.390905$ and $z_\textrm{\scriptsize red}=4.391265$, with a fixed $b=8$\,\kms. For the \CIV\ and \SiIV\ doublets, the transition providing the lowest column density upper limit was used in our photoionisation analysis (\Sref{s:results}). This is indicated by a ``*'' symbol. Note that for these doublets, our approach is to obtain a 2$\sigma$ upper limit on the column density by assuming that the total absorption seen at the nominated redshift is real. The last row provides the different metallicity upper limits inferred from the photoionisation analysis described in \Sref{s:results} using the lower and upper \NHI\ limits indicated.}
\label{t:TEST}
\begin{tabular}{c|c|c|c|c|c|c}
\hline
 Ion & \multicolumn{6}{c}{$\log_{10} (N/\mathrm{cm^{-2}})$} \\
     & \multicolumn{2}{c}{$\zab=4.391085$} & \multicolumn{2}{c}{$z_\textrm{\scriptsize blue}=4.390905$} & \multicolumn{2}{c}{$z_\textrm{\scriptsize red}=4.391265$} \\
 \hline
  \tran{Si}{ii}{1260} &\multicolumn{2}{c|}{  $<10.66       $}&\multicolumn{2}{c|}{  $<10.63    $}&\multicolumn{2}{c}{  $<10.62     $}  \\
  \tran{Si}{iv}{1393} &\multicolumn{2}{c|}{  $<12.04^*     $}&\multicolumn{2}{c|}{  $<12.06^*  $}&\multicolumn{2}{c}{  $<12.07^*   $}  \\
  \tran{Si}{iv}{1402} &\multicolumn{2}{c|}{  $<12.29       $}&\multicolumn{2}{c|}{  $<12.70    $}&\multicolumn{2}{c}{  $<12.49     $}  \\
  \tran{C}{ii}{1334}  &\multicolumn{2}{c|}{  $<11.96       $}&\multicolumn{2}{c|}{  $<11.89    $}&\multicolumn{2}{c}{  $<11.95     $}  \\
  \tran{C}{iv}{1548}  &\multicolumn{2}{c|}{  $<12.38^*     $}&\multicolumn{2}{c|}{  $<12.55^*  $}&\multicolumn{2}{c}{  $<12.82     $}  \\
  \tran{C}{iv}{1550}  &\multicolumn{2}{c|}{  $<12.76       $}&\multicolumn{2}{c|}{  $<12.70    $}&\multicolumn{2}{c}{  $<12.66^*   $}  \\
  \tran{O}{i}{1302}   &\multicolumn{2}{c|}{  $<12.00       $}&\multicolumn{2}{c|}{  $<12.00    $}&\multicolumn{2}{c}{  $<12.04     $}  \\
 \hline
 \lNHI & $17.9 \pm 0.1$  & $18.3 \pm 0.1$  & $17.9 \pm 0.1$  & $18.35 \pm 0.1 $  & $17.9 \pm 0.1$  & $18.20 \pm 0.1 $ \\
 \lmetal & $<-4.14$  & $<-4.31$ & $<-4.16$  & $<-4.36$  & $<-4.16$  & $<-4.31$ \\
\hline
\end{tabular}
\end{center}
\vspace{-0.5cm}
\end{table*}

The highly ionised transitions in \Fref{f:metals} all show some possible absorption signatures. However, for \ion{Si}{iv} the $\lambda$1393 transition falls at the extreme blue edge of the bluest extracted echelle order on the middle chip of HIRES's CCD mosaic. The decline in flux from $+$30 to $-20$\,\kms\ is likely to be an artefact of the flux extraction process and not likely to be real absorption. Indeed, somewhat different spectral shapes for this artefact are seen in the four different exposures contributing to the final spectrum in this range, further indicating its spurious origin. Nevertheless, this still provides the strongest 2$\sigma$ upper limit on the \ion{Si}{iv} column density because the $\lambda$1402 transition falls on the red edge of the \tran{Fe}{ii}{1608} transition of an unrelated absorption system at $\zab=3.697$; this is responsible for the deep absorption seen around $-20$\,\kms\ and, most likely, the much small amount of absorption at $v\approx0$\,\kms. To obtain a conservative upper limit on the \ion{Si}{iv} column density, we treat the absorption in each transition as if it were really \ion{Si}{iv}, and derive the values listed in \Tref{t:TEST}. However, it is likely that these greatly overestimate the true \ion{Si}{iv} column densities and we test how this affects our metallicity estimates by removing the \ion{Si}{iv} upper limit in \Sref{s:results}.

Both of the \ion{C}{iv} doublet transitions fall in regions with many telluric absorption lines. While it is possible that the broad trough of absorption bluewards of \tran{C}{iv}{1548} is partially caused by \ion{C}{iv} in LLS1723, its total column density is limited by the corresponding region in \tran{C}{iv}{1550}. That region in \tran{C}{iv}{1550} has several clear sky absorption features as well which further reduces the possible contribution of real \ion{C}{iv} absorption in both transitions. There is a strong telluric feature at 10\,\kms\ in \tran{C}{iv}{1548} but no strong absorption is seen at corresponding velocities in \tran{C}{iv}{1550}. As with \ion{Si}{iv}, our approach is to obtain a 2$\sigma$ upper limit on the \ion{C}{iv} column density by assuming that the total absorption seen at $v\approx0$\,\kms\ is really due to \ion{C}{iv} in LLS1723. These values are listed in \Tref{t:TEST}. We note that treating the broad absorption trough bluewards of \tran{C}{iv}{1548} as real absorption results in a total \ion{C}{iv} column density very similar to the upper limit listed in \Tref{t:TEST}. Therefore, we treat the \ion{C}{iv} column density as an upper limit for our main results but also test how treating it as a detection affects them in our photoionisation modelling (\Sref{s:results}).

\Fref{f:metals} shows a Voigt profile (red line) for each transition which represents the corresponding column density upper limit in \Tref{t:TEST} plotted at $\zab=4.391085$ and with a Doppler $b$ parameter of 8\,\kms. These visually demonstrate the suitability of the upper limits derived above. They also help to illustrate how the upper limits would change if LLS1723 were at slightly lower or higher redshift. In \Tref{t:TEST} we show how these limits change given the $\pm$10\,\kms\ redshift uncertainty discussed in \Sref{s:hydro} (see \Fref{f:lymanseries}), i.e.\ shifting the redshift bluewards to $z_\textrm{\scriptsize blue}=4.390905$ and redwards to $z_\textrm{\scriptsize red}=4.391265$. As expected from the discussion above, the low-ion column densities are robust to this redshift uncertainty, while the \tran{Si}{iv}{1402} and \ion{C}{iv} doublet transition column densities are more sensitive to it. However, the final \ion{Si}{iv} column density used in our photoionisation modelling (\Sref{s:results}) is entirely derived from \tran{Si}{iv}{1393} which changes little ($<$0.04\,dex) at these alternative redshifts, and the maximum change in the \ion{C}{iv} column density is 0.3\,dex. We use these alternative column density upper limits, in combination with the corresponding changes to the \lNHI\ upper limit at the same redshifts (18.35 and 18.20, respectively; see \Sref{s:hydro}), to check how the metallicity results change in \Sref{s:results}.

\section{Photoionisation model results}\label{s:results}

Unlike the much more hydrogen rich and self-shielded damped Lyman alpha systems (DLAs), LLSs are mostly ionised and the detected metal atoms/ions do not account for the entire metal column densities. Therefore, we used the photoionisation simulation software, \textsc{Cloudy} (version 13.03)\footnote{See \urlstyle{rm}\url{https://www.nublado.org/}.} \citep{2013RMxAA..49..137F}, to characterise the ionisation state of LLS1723 and, ultimately, infer its metallicity, $Z/Z_\odot$. Our \textsc{cloudy} modelling approach followed that of \citet{2015MNRAS.446...18C}: the gas cloud is modelled as a single phase slab, with a constant density and illuminated on one side by the UV background of \citet[hereafter HM12]{2012ApJ...746..125H} at $\zab=4.391085$. We generated a grid of photoionisation models covering hydrogen volume densities $-4.2 < \lnH < -1$, \HI\ column densities $17.5<\lNHI<18.5$, and metallicities $-6<\lmetal<-2$, all with respective steps of $0.2$\,dex. We assumed a solar abundance pattern according to the values from \cite{2009ARA&A..47..481A}.

We compared the upper limits of \Tref{t:TEST} to the values predicted by \textsc{cloudy}, as a function of \NHI, \nH\ and $Z/Z_\odot$. To do so, we used the same Markov Chain Monte Carlo (MCMC) sampling approach of \citet{2015MNRAS.446...18C} using the \textsc{emcee} code \citep{2013PASP..125..306F}. This approach seeks to maximise the likelihood, $\ln\mathcal{L}(\NHI,\nH,Z/Z_\odot)$, and determine the posterior distributions of $Z/Z_\odot$, \nH, and the ionisation parameter $U$ defined as $\Phi/\nH c$ with $c$ the speed of light and $\Phi$ the flux of ionising photons from the UV background.

Without any metal lines confidently detected, $U$ is not constrained in our model. We must therefore apply a prior on the density of the cloud (\nH) in the MCMC analysis. For LLSs, the range of densities found in previous, better constrained photoionisation models, was found to be $-3.5 < \lnH < -1.5$ \citep{2011Sci...334.1245F,2015ApJ...812...58C,2016MNRAS.455.4100F}, corresponding to a range of ionisation parameters $-4.5 < \log_{10} U < -2.5$ at the redshift of LLS1723. This range of densities represents the likely extremes that characterise LLSs arising in the intergalactic medium and circumgalactic environments, respectively: $\lnH\approx -3.5$ would represent an extremely diffuse intergalactic cloud which is unlikely even to form a LLS, whereas $\lnH\approx -1.5$ would characterise a very dense environment, similar to those in which DLAs arise \citep[e.g.][]{2015ApJ...800...12C,2016MNRAS.455.4100F}. We therefore adopt a flat prior over this density range in our MCMC analysis.

\begin{figure}
\begin{center}
\includegraphics[width=0.95\columnwidth]{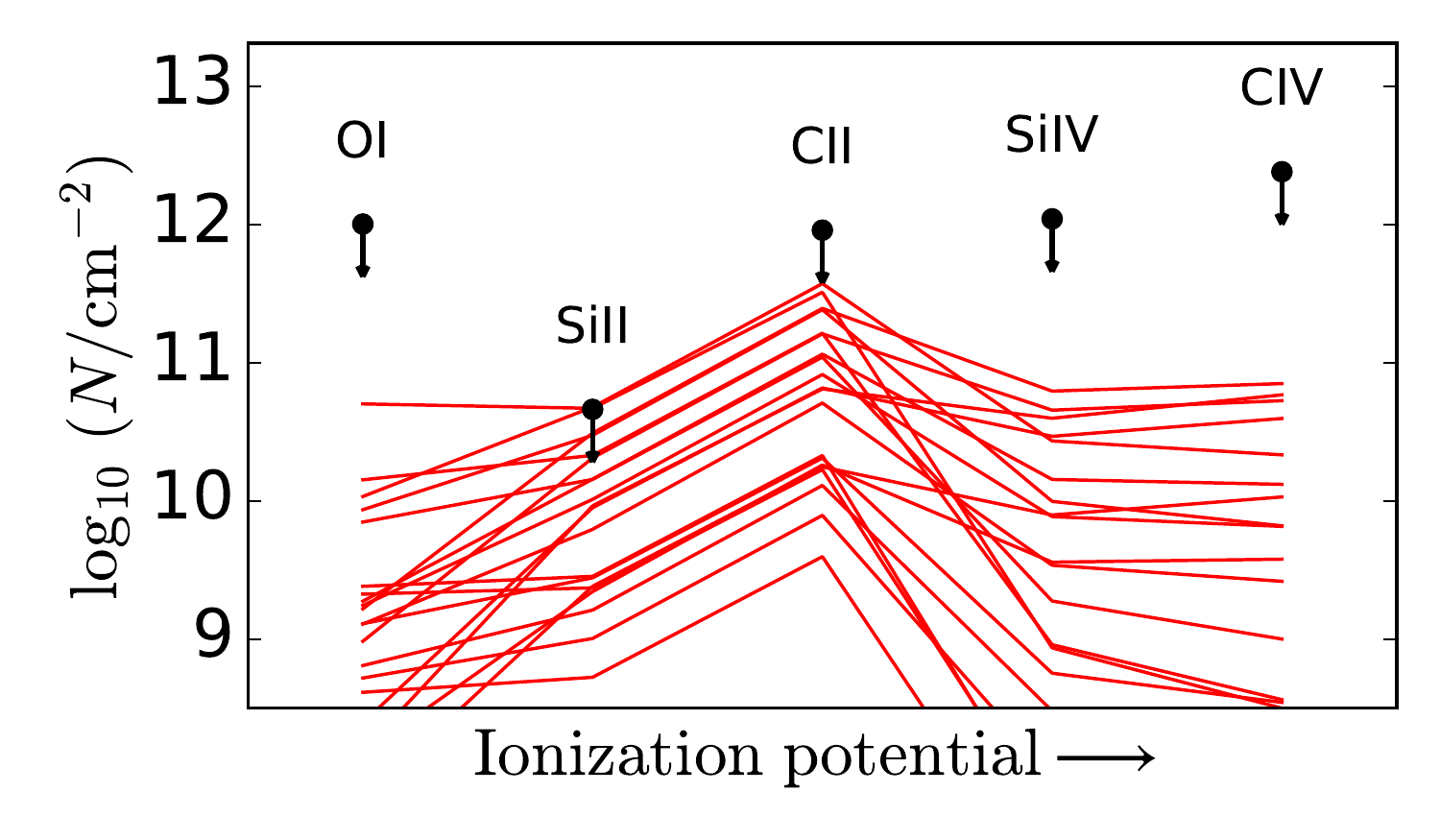}
\vspace{-0.4cm}
\caption{\label{f:models} Comparison between the ionic column density upper limits of \Tref{t:TEST} and the prediction of 20 \textsc{Cloudy} models drawn at random from the parameter distributions (solid red lines) for the fiducial model: $\zab=4.391085$ and $\lNHI = 17.9 \pm 0.1$.}
\end{center}
\vspace{-0.5cm}
\end{figure}

\Fref{f:models} shows the \textsc{cloudy} model predictions using the upper limits of \Tref{t:TEST} for our fiducial model. All the species listed are well-modelled, in the sense that their predicted \textsc{cloudy} column density values fall below our upper limit estimates: that is, our single phase model of LLS1723 appears a reasonable description of the data. For this highest metallicity model, the \ion{Si}{ii} and, to a lesser extent, \ion{C}{ii} upper limits prove to be the most important constraints on the total metal content of the absorber, whereas the MCMC sample is not strongly constrained by \ion{O}{i} or the high ions (\ion{Si}{iv} and \ion{C}{iv}). This aligns well with the reliability of the different metal-line upper limits derived from \Fref{f:metals}: the \tran{Si}{ii}{1260} transition occurs in a relatively ``clean'' region of the spectrum, not blended with strong telluric features or absorption features at other redshifts.

\begin{figure}
\begin{center}
\includegraphics[width=0.95\columnwidth]{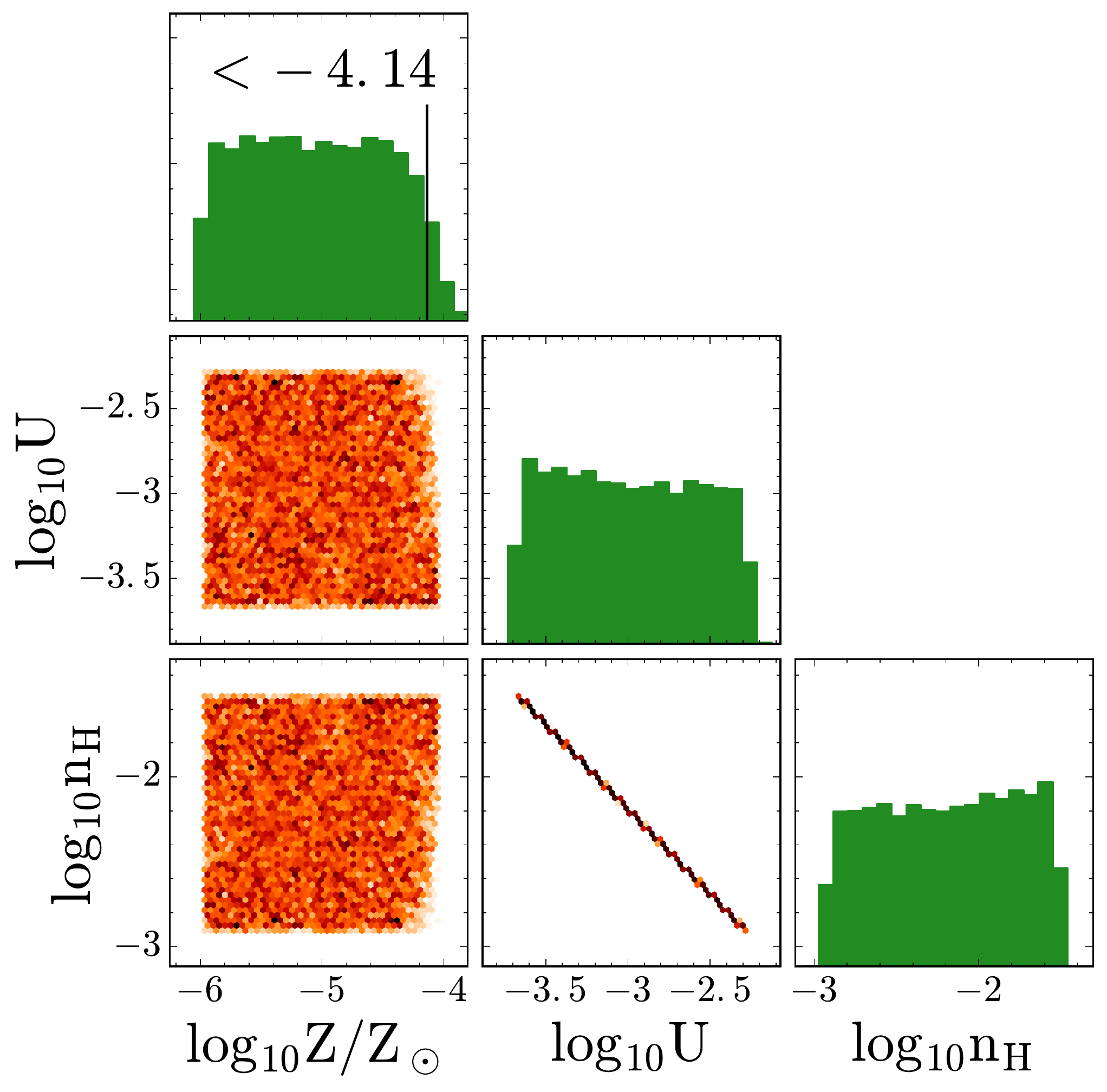}
\vspace{-0.4cm}
\caption{\label{f:triangle} Photoionisation model results for LLS1723 in the fiducial model: $\zab=4.391085$ and $\lNHI = 17.9 \pm 0.1$. The figure shows the MCMC distributions of parameter pairs involving metallicity ($Z/Z_\odot$), the ionisation parameter ($U$), and the volume density of hydrogen (\nH).}
\end{center}
\vspace{-0.5cm}
\end{figure}

\Fref{f:triangle} shows the distribution of $Z/Z_\odot$, \nH, and $U$ derived by the MCMC sampling algorithm using our fiducial model. This shows our main result: even using the lower limit on the \NHI\ column density ($\lNHI=17.9\pm0.1$) the 95\%-confidence upper limit on the metallicity of LLS1723 is $\lmetal < -4.14$. This is a remarkably low value; similarly low values have only been found serendipitously in two other LLSs \citep{2011Sci...334.1245F}. The physical implications of such a very low upper limit will be discussed below in \Sref{s:discussion}. The remainder of this section is dedicated to different consistency checks in order to test the robustness of this metallicity upper limit. \Fref{f:triangle} also demonstrates the relatively flat MCMC distributions for \nH\ and $U$ over the prior range set for \nH\ ($-3.5 < \lnH < -1.5$). Note that this large density range, together with the range of neutral fractions allowed by the photoionisation model ($\log_{10}x_\textsc{h}\approx-3$ to $-1.4$) implies a very large possible range of line-of-sight thicknesses for the cloud, $L \equiv \NHI/x_\textsc{h}\nH$, i.e.\ $\log_{10}(L/{\rm kpc})\approx 0.5$--5.2, which itself offers no insight into the origin of LLS1723.

To test the dependence of the metallicity estimate of LLS1723 on \NHI, we used the upper limit on its value, $\lNHI=18.3 \pm 0.1$, instead of the fiducial value ($\lNHI=17.9 \pm 0.1$), and kept the rest of the model parameters and metal-line column density upper limits the same. As the metallicity is expressed as a ratio over the total H column density, \NH, one should expect that increasing \NHI, and therefore \NH, would only decrease the metallicity upper limit for LLS1723. This is indeed the case: with the upper limit on \NHI\ the photoionisation model gives a 95\%-confidence upper limit on the metallicity of $\lmetal < -4.31$.

The $\pm$10\,\kms\ redshift uncertainty discussed in \Sref{s:hydro} introduced two new sets of column density upper limits, as listed in \Tref{t:TEST}. The low-ion column densities are robust to the redshift uncertainty, and the \tran{Si}{iv}{1402} and \ion{C}{iv} doublet column densities changed only by a small amount. As demonstrated in \Fref{f:models}, these high ions do not constrain the metallicity in our models very strongly. Therefore, we should expect the metallicity upper limit of LLS1723 to be robust to the changes in column density upper limits from the uncertainty in redshift. For the \NHI\ lower-limit of $\lNHI=17.9 \pm 0.1$, we obtained a 95\%-confidence metallicity upper limit of $\lmetal_\textrm{\scriptsize blue} < -4.16$ and $\lmetal_\textrm{\scriptsize red} < -4.16$ for the blue and red redshift limits in \Tref{t:TEST}. As discussed in \Sref{s:hydro}, the upper limit on \NHI\ changes for these two different redshifts. With $\lNHI=18.35 \pm 0.1$ at the blue limit, the metallicity obtained from the photoionisation model is $\lmetal_\textrm{\scriptsize blue} < -4.36$; with $\lNHI=18.20 \pm 0.1$ at the red limit, the metallicity obtained is $\lmetal_\textrm{\scriptsize red} < -4.31$. That is, in all these alternative models, the metallicity of LLS1723 is lower than in our fiducial model.

In \Sref{s:metals}, we outlined the possibility that \ion{C}{iv} was marginally detected in the broad trough of absorption seen bluewards of $v=0$\,\kms\ in \Fref{f:metals}. In that case, the total column density of \ion{C}{iv} was very similar to the fiducial upper limit provided in \Tref{t:TEST}. Using this value as a detection, rather than an upper limit, the photoionisation model returns a metallicity measurement (not upper limit) for LLS1723 of $\lmetal=-4.04 \pm 0.26$ (1$\sigma$ uncertainty). While this is slightly higher than the upper limit for our fiducial model, is does not alter our interpretation that LLS1723 is a very metal poor absorber. However, in this case our photoionisation model produces a mis-match between the predicted column densities of \ion{Si}{ii} and \ion{C}{iv}: the former is overpredicted by $\ga 0.2$\,dex while the later is underpredicted by $\ga 0.5$\,dex. The other ions did not strongly constrain these models, just as in the fiducial case. This mis-match indicates that, if the putative \ion{C}{iv} absorption is real, then a multiphase model for LLS1723 is required. Here, the low ions would be associated with the most neutral phase of the absorber, containing most of the \HI, and the high ions would represent a separate, more ionised phase. This follows the trend observed in large LLS samples that the low ions better match the \HI\ velocity structure \citep[e.g.][]{2013ApJ...770..138L,2016ApJ...831...95W}. This implies that our fiducial model, whose metallicity is constrained by the low ions, still provides a reliable metallicity upper limit. A simple consistency check demonstrates this clearly: removing the \ion{C}{iv} and \ion{Si}{iv} upper limits completely from the MCMC analysis results in an almost unchanged metallicity upper limit, i.e.\ $\lmetal < -4.12$.

Finally, we checked the impact of the shape of the UV background on our metallicity upper limit for LLS1723. Our fiducial model assumes the HM12 background, which is significantly softer at $\zab = 4.391085$\ than the previous version implemented in \textsc{cloudy}, often referred to as HM05, which is a revised version of that originally published by \citet{1996ApJ...461...20H}. The effect of these different backgrounds on LLS metallicities derived from \textsc{cloudy} photoionisation models has been extensively discussed \citep[e.g.][]{2013ApJ...770..138L,2014ApJ...792....8W,2015MNRAS.446...18C,2016ApJ...831...95W,2017ApJ...837..169P}. For instance, \citet{2016ApJ...831...95W} found that the HM12 background at $z<1$, where it is harder than the HM05 background, results in a higher metallicity than the HM05 background by 0.3\,dex on average; i.e.\ the harder background yields a higher metallicity. It is therefore important to consider how changing the UV background shape affects our metallicity upper limit. Nevertheless, with no metal-line detections, and the metallicity upper limit constrained mostly by the low ions, it is unlikely that this will impact our results significantly. As a first demonstration of this, we computed new \textsc{cloudy} model grids, in the same parameter space as our fiducial grids, using the HM05 background at the redshift of LLS1723, and derived a new metallicity upper limit using the same MCMC approach. There was no substantial change from our fiducial result: $\lmetal_\textrm{\scriptsize HM05} < -4.15$ at 95\% confidence. We also tested this by introducing a variable slope, $\alpha_\text{UV}$, to the HM12 UV background following \citet{2015MNRAS.446...18C}: the HM12 background is reproduced with $\alpha_\text{UV}\equiv 0$. If we allow $\alpha_\text{UV}$ to be determined by the MCMC process, a flat posterior probability distribution is obtained for $\alpha_\text{UV}$ in the range between $-2.5$ (very soft background) and $+1.5$ (very hard background). Even with this extra degree of freedom, the metallicity upper limit is again essentially unchanged from our fiducial model: $\lmetal < -4.13$ with 95\% confidence.
% "The important distinction between the 2005 revision of Haardt & Madau (1996, hereafter HM05), used in L13, and Haardt & Madau (2012, hereafter HM12) is the greatly reduced escape fraction of radiation from galaxies to match high-redshift data, leading to a harder UVB spectrum" 

\section{Discussion}\label{s:discussion}

Our result is that the non-detection of metals in LLS1723 implies a 95\%-confidence metallicity upper limit of $\lmetal < -4.14$. As discussed in \Sref{s:results}, this is likely a conservative upper limit because our neutral hydrogen column density estimate, $\NHI=17.9 \pm 0.1$, stems from the assumption that there is no residual flux below the Lyman limit (see \Fref{f:grid}); using the higher \NHI\ allowed by the \lya\ line alone would reduce the metallicity estimate. This very low metallicity raises questions about the origin of LLS1723 and we discuss the possibilities below.

\begin{figure}
\begin{center}
\includegraphics[width=0.95\columnwidth]{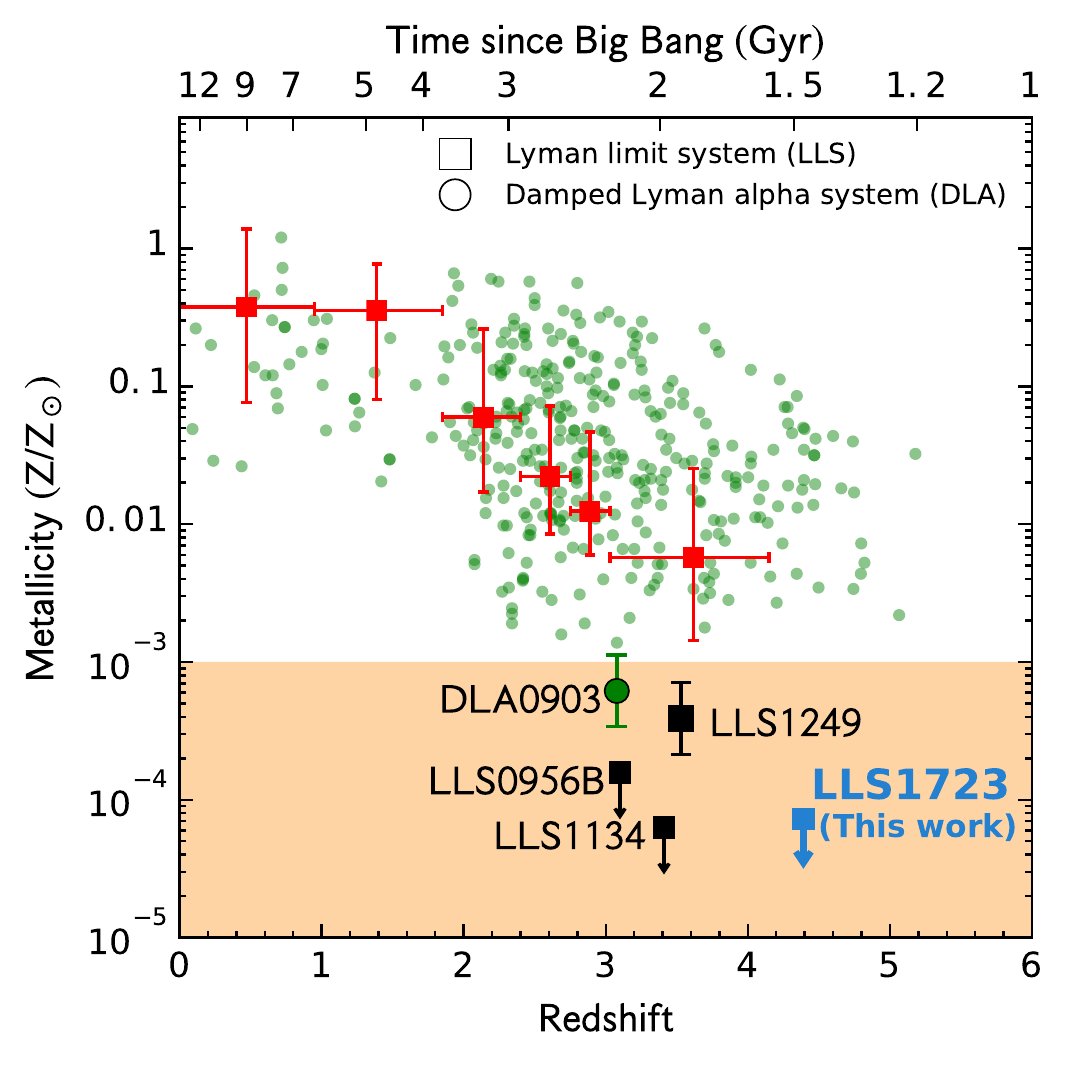}
\vspace{-0.4cm}
\end{center}
\caption{Summary of the metallicity distribution of LLSs (squares) and DLAs (circles) in the literature. Upper limits are indicated by arrows. The metallicity upper limit (95\% confidence) for LLS1723 from this work is highlighted in blue. Upper limits from \protect\citet{2011Sci...334.1245F} for two apparently metal-free LLSs, LLS0958B and LLS1134, are shown in black. The lowest metallicity measurement for a LLS from \protect\citet{2016MNRAS.457L..44C}, LLS1249, is shown in black with its associated 1$\sigma$ error bar. The red squares and error bars represent the LLS sample of \protect\citet{2016MNRAS.455.4100F} and show the median and 25--75\% range of the composite posterior probability density function in redshift bins containing at least 25 LLSs each. The green circles are DLAs from \protect\citet{2011MNRAS.417.1534C}, \protect\citet{2012ApJ...755...89R} and \protect\cite{2013MNRAS.435..482J}. The lowest metallicity measurement for a DLA from \protect\citet{2017MNRAS.467..802C}, DLA0903, is shown in green with its associated 1$\sigma$ error bar. The shaded orange region shows the expected metallicity range for gas enriched by PopIII supernovae from the simulations of \protect\citet{2012ApJ...745...50W}.}
\label{f:LLS}
\vspace{-0.1cm}
\end{figure}

\Fref{f:LLS} summarises the metallicity distribution of LLSs and DLAs in the literature. The LLS distribution has been measured by recent, large-scale studies, with 157 measurements at $z \sim 1.8-$4.4 from HIRES, ESI and MIKE by \citet{2015ApJS..221....2P} and \citet{2016MNRAS.455.4100F}, and a further 31 at $2.3 < z < 3.3$ observed with HIRES by \citet{2016ApJ...833..283L}. These appear to form a broad, unimodal distribution of metallicities, with the bulk of the population at $z>2$ having $\lmetal\approx-2$. The metalicity probability density function for all LLSs derived from the photoionisation analysis of \citet{2016MNRAS.455.4100F} implies that approximately 10\% of LLSs at $z\sim2.5$--3.5 have $\lmetal < -3$. However, only a single\footnote{\citet{2016ApJ...833..283L} reported a metallicity of $\lmetal\approx-3.4$ for another LLS, LLS0958A at $\zab=3.223$ towards the quasar SDSS J095852.19$+$120245.0. However, this estimate does not include the full velocity structure of observed metal absorption. We find that including all the metal absorption increases the inferred metallicity to $\lmetal\approx-2.8$, which is above our nominal threshold of $-3$ for near-pristine LLSs (Robert et al., in preparation). Therefore, we have not included this LLS in \Fref{f:LLS}. This LLS is discussed further at the end of this section.} LLS in which metals have been detected has been found to have a detailed metallicity measurement below $\lmetal=-3$, i.e.\ LLS1249 at $z\approx3.5$ illustrated in  \Fref{f:LLS} \citep{2016MNRAS.457L..44C}. In this sense, it may be surprising that three apparently metal free LLSs have been discovered -- LLS0958B and LLS1134 by \citet{2011Sci...334.1245F}, and LLS1723 in this work -- which have metallicities $\lmetal<-3.8$. Of course, our target selection and observing campaign were specifically designed to identify very low metallicity LLSs. This was clearly successful. However, as discussed in \Sref{s:observations}, our target selection drew on a sample of effectively $\sim$191 LLSs without strong metallicity biases. Therefore, including the other two apparently metal free cases, which were serendipitously discovered, it is clear that LLSs at $3 \la \zab \la 4.5$ with metallicity $\lmetal<-4$ are rare, but not extremely so: $\sim3/191 \sim 1.6$\% of the population. That is, it is currently unclear whether they are simply the very low metallicity tail of a unimodal LLS distribution, or whether they constitute a second mode. In other words, if the LLS metallicity distribution is unimodal, it may be that all LLSs have the same origin, with the common assumption that they arise in the circumgalactic medium (CGM). On the other hand, very metal-poor LLSs constituting a second mode of the metallicity distribution, would mean that they arise in a different environment: for instance, the intergalactic medium (IGM). The remainder of this section is then dedicated to investigate this question.

In the context of LLSs arising in the CGM, their population may include gas ejected from galaxies, cold accretion streams or other, virialised gas located in the CGM. With such a low metallicity, LLS1723 is unlikely to be outflowing gas or virialised, well-mixed CGM gas: these CGM components would be highly enriched with metals from supernovae in the host galaxy. Indeed, the much higher column densities of DLAs make them very likely to arise in these CGM components (and not, for example, the IGM), and the lowest-metallicity DLA known -- the $z\approx3.1$ absorber discovered by \citet{2017MNRAS.467..802C} (included in \Fref{f:models}) -- has $\lmetal\approx-3.2$, considerably higher than that our upper limit for LLS1723. This leaves the cold stream component as a likely origin for a LLS like LLS1723 in a circumgalactic environment. This agrees with a prediction of cosmological simulations: the streams of cold gas accreting into galactic haloes have column densities in the LLS range \citep[e.g.][]{2011MNRAS.418.1796F,2011MNRAS.412L.118F,2012MNRAS.421.2809V}, so we expect some LLSs to be very low metallicity or, possibly, completely pristine gas being polluted for the first time by CGM gas at $z\sim3$--4.5. Therefore, LLS1723 may indeed be part of a cold accretion stream. We should however note that modelling the CGM in a non-idealised context is difficult, despite the advances in numerical simulation over the years. The CGM in such works is usually under-resolved, and this could impact the ability to reproduce its properties infered through observations. For instance, \citet{2018arXiv180804369V} found that the covering fraction of LLSs increases with the resolution of the CGM: from 8\%\ to 30\%\ within 150 kpc from the galaxy centre.

On the other hand, the current small sample of very metal-poor LLSs might instead arise in a different environment: the IGM. There are several reasons to expect this scenario. The first is the redshift evolution of the number of LLSs per unit redshift, $l(z)$. In \citet{2013ApJ...775...78F}, the evolution of $l(z)$ at $z>3.5$ can not be explained with only a contribution from the CGM, i.e.\ a contribution from the IGM is speculated. Another motivation for the intergalactic scenario comes from the lowest-metallicity LLS with detected metals, LLS1249: on the basis of the [C/Si] ratio \citep[i.e.\ compared to metal-free nucleosynthesis calculations, e.g.][]{2002ApJ...567..532H} and metal-line shifts from the hydrogen lines, \citet{2016MNRAS.457L..44C} argued that LLS1249 may be an intergalactic remnant of a PopIII explosion, unpolluted by subsequent generations of stars. However, with only two detected metal species and metallicity $\lmetal\approx-3.4$, the possibility that this system is circumgalactic, and mainly polluted by PopII/I stars, remains. Finally, previous studies of the chemical abundance of the IGM using the \lya\ forest \citep[e.g.][]{2001ApJ...561..521A,2003ApJ...596..768S,2004ApJ...602...38A,2004ApJ...606...92S,2008ApJ...689..851A,2011ApJ...738..159S} suggest that a large fraction of the IGM at high redshift ($z\ga2$) is metal-poor. For instance, \citet{2011ApJ...738..159S} estimated that $\sim50\%$\ of the \lya\ forest has $\lmetal\leq-3.6$ at $z\sim4.3$. Despite LLSs having a higher \HI\ column density than most features in the \lya\ forest (i.e.\ $\lNHI \le 15.5$), the fact that LLS1723's metallicity is typical of the IGM is at least consistent with an IGM origin.

Therefore, an intergalactic environment for the apparently metal free LLSs (LLS0958B, LLS1134, LLS1723) is particularly interesting to consider: in principle, their extremely low metallicities may imply they are either (i) completely metal free gas clouds, (ii) remnants of PopIII explosions, or (iii) clouds polluted by PopII/I debris at extremely low levels. Given the lack of metal abundance information in such systems, it is difficult to definitively distinguish between these possibilities. However, existing numerical simulations can provide a guide to their relative likelihood, and we discuss these below.

Apparently metal free clouds like LLS1723 may be completely pristine, intergalactic gas -- surviving vestiges of the early universe that have never entered a large enough overdensity to be polluted by stellar debris. Cosmological simulations of structure formation, including pollution from PopIII stars, feature a very patchy and inhomogeneous metal enrichment \citep[e.g.][]{2007MNRAS.382..945T, 2010MNRAS.407.1003M, 2012ApJ...745...50W, 2018MNRAS.475.4396J}. Highly metal-enriched regions and pristine regions coexist in the simulation volume, but the latter only survive unpolluted in more isolated, low density regions, i.e.\ the intergalactic medium. These regions still exist, and are able to host PopIII star formation events, even at the end of the simulation; e.g.\ down to $z \sim 2.5$ in \citet{2007MNRAS.382..945T}. The very low metallicity upper limit in LLS1723, and the rarity of such clouds ($\sim 1.6$\% from our estimate above), are clearly compatible with this scenario.

However, it is also possible that PopIII stars could have enriched intergalactic LLSs, but to levels below $\lmetal\approx-4$. Clearly, the enrichment level will depend on many factors, including the mass of the progenitor star, its explosion energy and the mechanisms and timescales for mixing the ejecta with the surrounding pristine gas. Therefore, understanding the detailed fate of PopIII remnants is complex, and numerical simulations require very high spatial resolution (below the parsec level) to accurately model it; this limits the parameter space that can be explored. Nevertheless, in existing simulations, the mass of the progenitor star, which governs the amount of metal ejected, appears to largely determine whether enrichment below $\lmetal \la 4$ is possible. For instance, in the simulation by \citet{2016MNRAS.463.3354R}, which has a minimum cell size of $0.02$\,pc, a single 60\,\msun\ PopIII star explodes as a low-energy supernova ($\approx 10^{51}$\,erg) at $z\approx19$ and enriches its low-mass host halo ($\approx 10^6$\msun) to a metallicity of 2--$5 \times 10^{-4}Z_\odot$, not far above the metallicity limit for LLS1723. At the other end of the PopIII mass scale, individual pair-instability supernovae at $z=13$--16, with masses in the range 140--260\,\msun, enrich their host halos to metallicities above $Z \approx 10^{-3}Z_\odot$ in the simulation by \citet{2012ApJ...745...50W}; similar results were found by \citet{2010MNRAS.407.1003M}. Therefore, LLS1723 may have been enriched, but to undetectable levels, by a low-mass PopIII star, but pollution by high-mass PopIII stars appears less likely.

It also appears unlikely that PopII/I stars would enrich an intergalactic, pristine cloud to metallicities below $\lmetal\approx-4$. Such stars will form from PopIII-enriched gas which, as described above, likely already has $Z \ga 10^{-4}Z_\odot$, before substantial PopII star formation occurs \citep[e.g.][]{2010MNRAS.407.1003M}, and so the added enrichment from PopII/I explosions will significantly exceed the metallicity upper limit found for LLS1723. Therefore, if LLS1723 is intergalactic, most simulations indicate that we should not expect it to have been enriched by PopII/I stars.

Given the above considerations from simulations, it appears most likely that LLS1723 is either:
\begin{enumerate}
\item in an intergalactic environment, and either completely pristine gas (no stellar pollution) or has been polluted by supernovae from low-mass PopIII stars only at very high redshifts (i.e.\ $z\ga10$); or
\item cold stream gas encountering the circumgalacitc medium for the first time at the observed redshift ($z\approx4.4$). In this case, the gas may have the same enrichment history, or lack of it, as in the intergalactic case, but may now undergoing pollution by the $z=4.4$ CGM as well.
\end{enumerate}
Understanding the origin of the very low metallicity LLSs will, therefore, partly rely on distinguishing between these intergalactic and circumgalacitc possibilities. A simple test between these scenarios would be to map the \lya-emitting galaxy environment around systems like LLS1723 using optical integral field spectroscopy: if the system is intergalactic then no galaxies will be found close to the quasar line of sight at the system's redshift; if the system is circumgalactic then, depending on the flux limits of the observations, a galaxy (or galaxies) will be detected nearby.

Before the advent of optical integral field spectrographs such as KCWI \citep{2018ApJ...864...93M} on the Keck II telescope and MUSE \citep{2010SPIE.7735E..08B} on the VLT, such a test was time-consuming and difficult, requiring multiband imaging (to pre-select galaxies near the system's redshift) followed by spectroscopy of candidate galaxies. However, using MUSE, \citet{2016MNRAS.462.1978F} demonstrated that relatively short ($\sim$5\,hrs) integral field observations in a single pointing was sufficient to map the \lya-emitting galaxy environment down to luminosities $L_{\lya} \ge 3\times10^{41}$\,erg\,s$^{-1}$ ($\la0.1L^*_{\lya}$) in a $\sim$160\,kpc radius around two low metallicity $\zab\approx3$ absorbers towards the quasar SDSS J095852.19$+$120245.0. One of these, LLS0958A, showed no nearby galaxies, potentially indicating an intergalactic environment. Interestingly, this LLS has weak-but-detected metal lines \citep[][see footnote 4]{2016ApJ...833..283L}. Given our discussion above, this may indicate it has a PopIII origin; however, as with LLS1249 \citep{2016MNRAS.457L..44C}, only the [C/Si] ratio could be determined, so this origin cannot be confirmed by comparison with nucleosynthetic yield models. The other LLS studied by \citet{2016MNRAS.462.1978F} is LLS0958B, one of the apparently metal free systems (like LLS1723) shown in \Fref{f:LLS}. Five \lya\ emitters at the redshift of LLS0958B were detected, three of which appear aligned in projection and may indicate a filamentary intergalactic structure. That is, LLS0958B may indeed correspond to case (ii) above: cold stream gas entering a circumgalactic environment for the first time. Even from just these two examples, it is clear that we should not expect just one type of galaxy environment for very low metallicity LLSs, and they underscore how galaxy mapping around such systems can assist in understanding their origin.

\section{Conclusions}\label{s:conclusion}

In this work, we reported the discovery of a new, apparently metal-free Lyman limit system, LLS1723, located at redshift $\zab=4.391$ towards the quasar PSS1723$+$2243 at $\zem=4.515$, based on observations with the HIRES spectrograph. LLS1723 is part of our dedicated search of very metal-poor Lyman limit systems in order to study their origins. LLS1723 exhibited no detected metal absorption lines in a previous ESI spectrum from \citet{2015ApJS..221....2P}, suggesting a very low metallicity, $\lmetal < -3$. We did not convincingly detect any metal lines in HIRES spectrum, confirming this initial selection and providing an order-of-magnitude lower metallicity upper limit.

Our conservative upper limit on the metallicity of LLS1723 is characterized by a hydrogen column density of $\lNHI=17.9 \pm 0.1$, defined by the lack of observed flux below its Lyman limit. Upper limits on the strongest metal line column densities were established with the apparent optical depth method, with that from \tran{Si}{ii}{1260} proving to constrain the metallicity most strongly. With the combined use of a grid of \textsc{cloudy} photoionisation models and MCMC sampling techniques, we derived the upper limit of $\lmetal < -4.14$ at 95\% confidence. A series of consistency checks confirmed that this upper limit is both robust and conservative. For instance, a possible detection of high ions in our HIRES spectrum of LLS1723 (specifically \ion{C}{iv}) leads to only a marginally higher metallicity estimate ($-4.04$), but also implies that a two-phase model of LLS1723 would be required to simultaneously match the low and highly ionised metal column densities. In such a two-phase model, the low-ionization phase would still have a metallicity constrained by \tran{Si}{ii}{1260}, as in our fiducial model, and so would have $\lmetal < -4.14$. The hydrogen column density of LLS1723 may also be as large as $\lNHI=18.3 \pm 0.1$, as limited by a flux transmission peak near its \lya\ absorption line. In that case, the metallicity upper limit would reduce to $-4.31$, which would be the lowest limit among the three apparently metal-free LLSs currently known.

Focussing on its very low metallicity, we discussed in \Sref{s:discussion} the possible origin scenarios for LLS1723, considering both circumgalactic and intergalactic environments. In the case of a circumgalactic origin for LLS1723, the most likely scenario appears to be cold-stream gas encountering the CGM for the first time at $z=4.4$. We considered very unlikely the possibility of LLS1723 being part of a galactic outflow or virialised, well-mixed circumgalactic gas. If instead LLS1723 arises in an intergalactic environment, it may well be a truly metal-free gas cloud, that has remained unpolluted since the Big Bang for 1.4\,Gyr. This appears compatible with the survival of rare pristine regions in numerical simulations, and would also explain the apparent (though clearly uncertain) rarity of potentially metal-free LLSs. Alternatively, LLS1723 may be an intergalactic gas cloud that has been enriched to a metallicity below our upper limit of $\lmetal < -4.14$. One possible way to have such a low level of metal pollution is via low-energy supernova of low-mass PopIII stars. Very massive PopIII supernova, especially very energetic pair-instability supernovae, would most likely enirch surrounding gas clouds to levels above $\lmetal \sim -3$. We also considered the enrichment of a pocket of intergalactic gas by PopII/I supernovae; however, this appears unlikely given that PopII/I stars would form in region already enriched above $\lmetal \sim -4$, and that their supernova ejecta would be unlikely to reach a pocket of intergalactic pristine gas. arguments for both cases relied on the interpretation of properties displayed by numerical simulations.

The interpretations of the origins for LLS1723 above are drawn from the general results of numerical simulations of the CGM and stellar nucleosynthetic enrichment of gas clouds. However, the possibility that LLS1723 and other very low-metallicity systems may arise in the IGM motivates detailed simulations that predict the frequency with which LLSs should arise in the IGM, not just in circumgalactic environments. Predictions for the relative cosmological abundance of intergalactic LLSs with different enrichment histories -- pristine, low-energy PopIII enrichment etc.\ -- would be invaluable for interpretting a sample of apparently metal-free and near-pristine LLSs. Observationally, mapping the galaxy distribution around such systems may assist in better understanding their origin. New optical integral field spectrographs, such as MUSE and KCWI, are already demonstrating this to be a promising approach \citep{2016MNRAS.462.1978F}. Nevertheless, the fact that these early results show quite very different galaxy distributions around two very low metallicity LLSs already shows that a systematic approach on a well-defined, statistical sample is likely to be required for reliable interpretations to be drawn.

Finally, this work has demonstrated that very metal poor LLSs can be discovered and studied in a targetted way, and opens the possibility for creating such a well-defined, statistical sample in future. This would help define the true metallicity distribution of LLSs at $z \geq 2$, which in recent surveys \citep[e.g.][]{2016MNRAS.455.4100F,2016ApJ...833..283L} appears unimodal, though very broad. This may be expected if LLSs share a common origin in the CGM. However, if a second population arises in the IGM -- apparently metal-free systems like LLS1723, and near-pristine systems like that in \citet{2016MNRAS.457L..44C} -- they would constitute a second mode that may occupy a separate metallicity range to the higher-metallicity LLSs arising in the CGM. Such a separation would assist comparison with the results of LLS simulations because the specific origin of each LLS may not need to be established individually and exhaustively; a statistical comparison may be sufficient to draw meaningful conclusions.

\section*{Acknowledgements}

\vspace{-0.2cm}
We thank Ryan Cooke for helping with part of the data reduction. PFR acknowledges supports through a Swinburne University Postgraduate Research Award (SUPRA) scholarship, and travel supports through the International Telescopes Support Office. MTM thanks the Australian Research Council for \textsl{Discovery Project} grant DP130100568 which supported this work. M.F. acknowledges support by the Science and Technology Facilities Council [grant number ST/P000541/1]. This project has received funding from the European Research Council (ERC) under the European Union's Horizon 2020 research and innovation programme (grant agreement No 757535). Our analysis made use of \textsc{astropy} \citep{2013A&A...558A..33A}, \textsc{matplotlib} \citep{Hunter2007}, and \textsc{barak} (\url{https://github.com/nhmc/Barak}). The data presented herein were obtained at the W.M. Keck Observatory, which is operated as a scientific partnership  among the California Institute of Technology, the University of California and the National Aeronautics and Space Administration. The Observatory was made possible by the generous financial support of the W.M. Keck Foundation. Australian access to the W. M. Keck Observatory has been made available through Astronomy Australia Limited via the Australian Government's National Collaborative Research Infrastructure Strategy, via the Department of Education and Training, and an Australian Government astronomy research infrastructure grant, via the Department of Industry, Innovation and Science. The authors wish to recognize and acknowledge the very significant cultural role and reverence that the summit of Maunakea has always had within the indigenous Hawaiian community. We are most fortunate to have the opportunity to conduct observations from this mountain. We are also grateful to the staff astronomers at Keck Observatory for their assistance with the observations. 

%%%%%%%%%%%%%%%%%%%%%%%%%%%%%%%%%%%%%%%%%%%%%%%%%%



  \bibliographystyle{mnras}
  \bibliography{biblio}{}

%%%%%%%%%%%%%%%%%%%%%%%%%%%%%%%%%%%%%%%%%%%%%%%%%%

%%%%%%%%%%%%%%%%% APPENDICES %%%%%%%%%%%%%%%%%%%%%

%\appendix



%%%%%%%%%%%%%%%%%%%%%%%%%%%%%%%%%%%%%%%%%%%%%%%%%%

%\section*{Supporting Information}\label{sec:supp}

%Additional Supporting Information may be found in the online version
%of this article:\vspace{-0.5em}\newline

%\noindent \textbf{Table/Figure reference or file name.} A short description (just a few words).\vspace{-0.1em}\newline
%\noindent \textbf{Table/Figure reference or file name.} A short description (just a few words).\vspace{-0.5em}\newline

%\noindent Please note: Oxford University Press are not responsible for the
%content or functionality of any supporting materials supplied by
%the authors. Any queries (other than missing material) should be
%directed to the corresponding author for the paper.

% Don't change these lines
\bsp	% typesetting comment
\label{lastpage}
\end{document}